\documentclass[twocolumn,showpacs,pra,aps]{revtex4}
\usepackage{graphicx}

\begin{document}

\title{Quantum entanglement with acousto-optic modulators: 2-photon beatings and Bell experiments with moving beamsplitters}

\author{Andr\'{e} Stefanov$^1$, Hugo Zbinden$^1$, Nicolas Gisin$^1$, and Antoine Suarez$^2$}
\address{$^1$Group of Applied Physics, University of Geneva, 1211 Geneva 4, Switzerland\\
$^2$Center for Quantum Philosophy, P.O.Box 304, 8044 Zurich,
Switzerland; suarez@leman.ch}
\date{\today}

\begin{abstract}

We present an experiment testing quantum correlations with
frequency shifted photons. We test Bell inequality with 2-photon
interferometry where we replace the beamsplitters by
acousto-optic modulators, which are equivalent to moving
beamsplitters. We measure the 2-photon beatings induced by the
frequency shifts, and we propose a cryptographic scheme in
relation. Finally, setting the experiment in a relativistic
configuration, we demonstrate that the quantum correlations are
not only independent of the distance but also of the time
ordering between the two single-photon measurements.
\end{abstract}

\pacs{03.65.Ta, 03.65.Ud, 03.30.+p, 03.67.Hk, 03.67.Dd, 42.65}

\maketitle

\section{Introduction}

Entanglement is a basic resource for quantum information
processing as well as for fundamental tests of quantum mechanics.
Several types of entanglement between photons have already been
demonstrated: polarization entanglement \cite{kwiat}, energy-time
entanglement \cite{Franson89, Brendel92, Tittel98} and time-bin
entanglement \cite{Brendel,Thew}, see \cite{WeihsTittel} for a
review. In this paper we present a setup based on energy-time
entanglement, where we add a frequency shift in one arm of each
interferometer. Experimentally the frequency shift is induced by
using acousto-optic modulators (AOM) in the interferometers
instead of standard beamsplitters.

The section II of this article is devoted to the effects of a
frequency shift on the time dependent coincidence rate in a
Franson-type Bell experiment. This effect is the equivalent for
2-photon interferences to the phenomenon of beatings for single
photon interferences. As the time needed to record interference
fringes can not be arbitrarily small, the measured visibility is
reduced in presence of beatings. When this measurement time is
only limited by energy resolution, there is a simple relation
between the visibility and the which-path information.
Experimentally we are far from accede to very short measurement
times, therefore we propose an indirect method to show the
beatings.

The section III presents the experimental setup in detail and the
technics used to overcome the difficulties due to the frequency
shifts. We have measured high visibility interference fringes when
the beatings are canceled, and we have also measured the beatings
frequency when it is not zero.

Since an AOM is equivalent to a moving beamsplitter, our setup can
be used to perform experiments with apparatuses in two different
relevant frames \cite{Stefanov}. In the conventional experiments
with all apparatuses at rest, there is only one relevant inertial
frame, i.e. one inertial frame of the massive pieces of the
apparatus (the laboratory frame) and, therefore, only one
possible time ordering: one of the photons is always measured
before the other (\emph{before-after} situation). Using two
relevant frames it is possible to create a \emph{before-before}
time ordering, in which each measuring device in its own inertial
frame analyzes the corresponding photon before the other, and an
\emph{after-after} time ordering, in which each measuring device
in its own inertial frame analyzes the corresponding photon after
the other \cite{Maudlin,asvs97}. Quantum Mechanics predicts
correlations independently of the time ordering, between the two
single-photon measurements. By contrast, Multisimultaneity
\cite{as00}, a recently proposed alternative theory, casts
nonlocality into a time ordered scheme and predicts disappearance
of nonlocal correlations with \emph{before-before} timing.
Therefore, experiments with AOMs allow us to test a most
important feature of quantum entanglement as it is the
independence of the time ordering. This is the subject of the
section IV.

\section{two photon beatings}

When two monochromatic waves of frequencies $\omega _{1}$ and
$\omega _{2}$ are combined the resulting wave exhibits two
frequencies, one at $\omega _{0}=\left( \omega _{1}+\omega
_{2}\right) /2$ and the other at $\delta \omega =\left( \omega
_{1}-\omega _{2}\right) /2.$ This is the well known phenomenon of
beatings. An application in the optical domain for classical
light field is heterodyne detection \cite{HP}.

\begin{figure}[h]
\includegraphics[width=8.43cm]{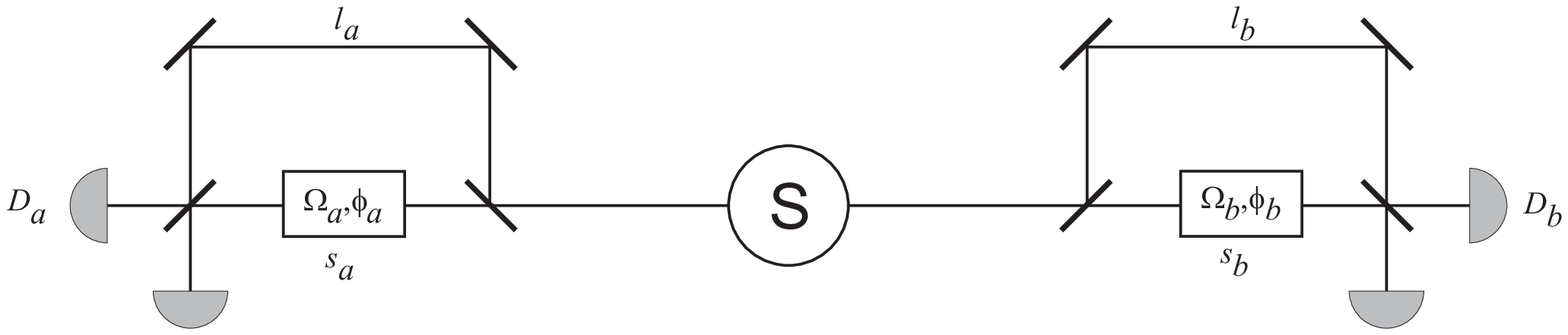}
\caption{Franson-type Bell experiment with frequency shift. $l_i$
and $s_i$ are the lengths of the long and short arms of
interferometer $i$, $\Omega _{i}$ and $\phi_i$ are the frequency
shift and phase shift in the short arm of interferometer $i$.}
\label{franson}
\end{figure}

Beatings can be seen as first order interferences in the time
domain. For second order interferences, the same equivalence can
be found. Consider the Franson-type configuration of figure
\ref{franson}; a source S emits energy-time entangled pairs of
photons. Each photon is sent to an unbalanced interferometer.
When both photons arrive in coincidence on the detectors it is
impossible to distinguish between both photons passing by the
short arms (\textit{ss}) or both passing by the long ones
(\textit{ll}) because the photons emission time is undeterminated.
Hence interference fringes appear when the phases $\phi_i$ are
changed. In our experiment, we consider not only phase changes in
each interferometers but also changes of the photons frequencies.
When both frequency shifts do not sum to 0 we will show that the
coincidence rate between two detectors changes periodically in
time. This is equivalent, for 2-photon interferences, to the
phenomenon of beatings in one photon interferences, therefore we
call it 2-photon beatings.

Following closely Franson's calculation \cite{Franson89}, the wave
function at detector $D_i$, $i\in\{a,b\}$, located at $r_i$, is
given by
\begin{equation}
\Psi \left( r_{i},t\right) =\frac{1}{2}\Psi _{0}\left( r_{i},t\right) +\frac{%
1}{2}e^{i\phi _{i}}\Psi _{0}^{\Omega _{i}}\left( r_{i},t-\Delta
t_{i}\right) \label{psi}
\end{equation}
where $\phi _{i}$ is the phase of interferometer $i$, $%
c\Delta t_{i}=\Delta l_{i}=$ $l_{i}-s_{i}$ is the path difference
between both arms of the interferometer and $\Omega _{i}$ the
frequency shift in the short arm of interferometer $i$. The
difference with Franson's original calculation is that we consider
an arbitrary frequency shift $\Omega_i$ in one arm of the
interferometers. We can expand the wave functions in the field
operators $c_{k}$
\begin{eqnarray}
\Psi _{0}\left( r_{i},t\right) &=&\sum_{k}c_{k}e^{i\left(
kr_{i}-\omega
_{i}t\right) }  \label{serie}\nonumber \\
\Psi _{0}^{\Omega_i }\left( r_{i},t\right)
&=&\sum_{k}c_{k}e^{i\left( kr_{i}-\left( \omega _{i}+\Omega
_{i}\right) t\right) }
\end{eqnarray}
where $\omega_i$ is the frequency of the photon in interferometer
$i$.
 The coincidence rate between 2 detectors is then given by
\begin{equation}
R=\left\langle 0\right| \Psi ^{+}\left( r_{a},t\right) \Psi
^{+}\left( r_{b},t\right) \Psi \left( r_{b},t\right) \Psi \left(
r_{a},t\right) \left| 0\right\rangle  \label{coin}
\end{equation}
Using (\ref{psi}) and (\ref{serie}) in (\ref{coin}) we find
\begin{equation}
R\sim\frac{1+\cos (\omega_0\Delta t+\Omega ^{0}\Delta t
+\phi_1+\phi_2 -\Omega ^{0}t)}{2} \label{Coincidences rate with
beatings}
\end{equation}
where $\Omega ^{0}=\Omega _{a}+\Omega _{b}$. This corresponds to
Franson's result, when $\Omega ^{0}=0$, otherwise the coincidence
probability is generally not constant in time.

When the entangled photons are created by downconversion
\cite{down}, we have to take into account the finite bandwidth of
the pump laser and of the photons. We find, assuming gaussian
spectral distributions, that the coincidence probability is given
by

\begin{equation}
R=\frac{1+\chi\cos (\omega_0\Delta t+\Omega ^{0}\Delta t
+\phi_1+\phi_2 -\Omega ^{0}t)}{2} \label{coincidence rate}
\end{equation}
with

\begin{eqnarray}
\chi =f(\frac{\Delta l_{b}}{c},\delta \omega _{0})f(\frac{\Delta
l_{a}-\Delta l_{b}}{c},\Delta )
 \label{Visibility}
\end{eqnarray}

where $f\left( x,y\right) =\exp \left(-\frac{1}{2}x^{2}y^{2}\right) $, $%
\omega_{0}$ is the central frequency of the pump laser, $%
\delta \omega _{0}$ is the pump bandwidth and $\Delta $ is the
photons bandwidth. Hence the visibility of the interference
fringes is reduced by a factor $\chi$. In absence of beatings
($\Omega^0=0$), $\chi$ is the maximal visibility that can be
measured. Equation (\ref{Visibility}) contains all the usual
conditions to see high visibility interference fringes; the
coherence length of the pump laser has to be greater than the
path difference in one interferometer, and the photons coherence
length has to be greater than the difference $\Delta l_{a}-\Delta
l_{b}$.

An application of entangled photons, apart from fundamental tests
of quantum mechanics, is quantum cryptography. Setups based on
polarization, energy-time or time-bin entanglement have been
proposed and realized, for a review see \cite{crypto}. One the
other hand only one photon schemes with frequency shifted photons
have been proposed \cite{Merolla99,Fainman}. In appendix A we
propose 2 different schemes with frequency shifted entangled
photons, which can be used to implement quantum cryptography,
although they are not actually of practical interest, due to
technology limitations.

\subsection{Frequency shift as quantum eraser}

If the beating frequency is not zero, the coincidence probability
changes in time, decreasing the visibility of the interference
fringes. According to Feynmann "principle" \cite{feynmann}, the
disappearance of the interference fringes would implies
accessibility, in principle, of information about which path the
photons took \cite{GHZ}. The frequency shift can be used to mark
the path, only if we have enough coincidence events such that the
time needed to experimentally estimate the coincidence
probability is smaller than the intrinsic uncertainty $\Delta t$
on the time measurement given by saturating the energy-time
uncertainty relation $\Delta E\Delta t= \hbar$. Otherwise
information about the path is lost due to imperfect experimental
devices.

We can quantify this information and the corresponding loss of
visibility. Contrary to \cite{schwindt} where the degree of
freedom used to mark the photon (i.e. polarization) is different
from the one where interferences are observed (spatial mode), we
use the frequency to mark the paths. This will affect the
interferences by creating beatings as we have shown before.
However due to the energy-time uncertainty relation there is
still a relation between interferences visibility and which-path
information.

If the time needed to measure the coincidence probability is
arbitrary small: $\Delta t=0$, the coincidence probability will
be given by $\frac{1+\cos \left( \Phi -\Omega^0 t\right) }{2}$,
according to (\ref{coincidence rate}) where we assume $\chi=1$ and
$\Phi=\omega\Delta t+\Omega ^{0}\Delta t +\phi_1+\phi_2$. For a
finite time resolution we have to integrate this expression over
a time distribution with a width $\Delta t$, for example we
consider a gaussian distribution:
\begin{eqnarray}
p &=&\int_{-\infty }^{\infty }\frac{1+\chi\cos \left( \Phi -\Omega^0 t\right) }{2%
}\frac{1}{\sqrt{2\pi }\Delta t}\exp \left(
-\frac{1}{2}\frac{t^{2}}{\Delta
t^{2}}\right) dt \nonumber\\
&=&\frac{1}{2}+\frac{1}{2}\cos \left( \Phi \right) \exp \left(
-\frac{1}{2} \left( \Omega^0 \right) ^{2}\Delta t^{2}\right)
\end{eqnarray}
The corresponding visibility is
\begin{eqnarray}
V\left( \Delta t\right) &=&\frac{p_{\max }-p_{\min }}{p_{\max }+p_{\min }}\nonumber\\
&=&\exp \left( -\frac{1}{2}\left( \Omega^0 \right) ^{2}\Delta
t^{2}\right) \label{VvsDt}
\end{eqnarray}
The which-path information is given by measuring the total energy
of the photons $\omega$. This is done with a resolution
$\Delta\omega$ related to the time resolution by the energy-time
uncertainty relation $\Delta\omega\Delta t\geq 2\pi$. We predict
that the photons would be detected in the $ll$ arm if we measure
$\omega<\omega_0+\Omega^0/2$, and in the $ss$ arm otherwise. The
which-path information $K$ is given by $K=2q-1$ \cite{Jaeger}
where $q$ is the probability of a correct prediction on the path.
With our strategy we have
\begin{eqnarray}
q&=&p(ll|\omega<\omega_0+\Omega^0/2)\nonumber\\
&=&1-\frac{1}{\sqrt{2\pi}\Delta\omega}\int_{\omega_0+\Omega^0/2}^{\infty}exp(-\frac{1}{2}\frac{(\omega-\omega_0)^2}{\Delta\omega^2})d\omega
\end{eqnarray}
Hence the information is given by
\begin{equation}
K=2erf(\frac{\Omega^0\Delta t}{4\sqrt{2}\pi})-1
\end{equation}
with $erf(x)$ being the error function. The extreme cases are
either a perfect distinction between $ss $ and $ll$ events which
requires that $\Delta E<<\hbar \Omega ^{0}$, but this implies a
measurement time $\Delta t>>1/\Omega ^{0}$, averaging to zero the
interferences; or, on the contrary, if the measuring time is short
enough to measure interference, then the energy resolution is to
poor to distinguish the paths. For the intermediate case we have
the known relation \cite{Jaeger} $V^2+K^2\leq1$. The equality is
not reached because of the prediction strategy is not optimal.

Let us emphasize that the preceding description does not rely on
quantum mechanics but more generally on wave theory. The quantum
nature appears when we assume that photons are quanta of light,
and in the fact that the photon pairs we consider can not be
described by classical local physics.

\subsection{Measurements of 2-photon beatings}

In our experiment, when the RF drivers are not sychronized, the
minimum value which can be set for $2\pi\Omega^{0}$ is $31.5$
KHz. This is too large to directly see the beatings by recording
the coincidence rate vs. time. A first possible method, is to
record the time of each coincidence event and reconstruct the
beatings from those datas. This requires a clock precise enough
over a long time. However this requires also that the coherence
of the beatings signal is much longer than the acquisition time,
so that the phases of the interferometers have to be kept stable
during that time. We also need to know precisely the frequency
$\Omega^{0}$, otherwise the analysis of the datas will be much
more complicated, although not impossible.

Instead of recording all the absolute times of arrival and
reconstruct the beatings we could only measure the time
difference between 2 successive coincidence counts and then
measure the distribution of those times.

The probability density $P(\Delta t)$ of having two coincidences
separated by a time $\Delta t$ can be computed in the following
way. Because the detection process is independent of time, except
for a dead-time of the detectors $\tau _{d}$, the conditional
probability of having a coincidence at time $t$ and another one in
time $t+\Delta t$, knowing that a photon pair reaches the
detectors at time $t$ and another one at time $t+\Delta t$ is
only dependent of the beatings signal.
\begin{eqnarray}
p\left( t,t+\Delta t|\gamma _{t},\gamma _{t+\Delta t}\right) = \nonumber\\
\newline
\left\{
\begin{array}{c}
{0\text{ if }\Delta t<\tau
_{d}} \\
{C\frac{1+V\cos \left( \Omega^0
t\right) }{2}\frac{1+V\cos \left( \Omega^0 \left( t+\Delta t\right) \right) }{2%
}\text{ if }\Delta t>\tau _{d}}
\end{array}
\right.
\end{eqnarray}

where $V$ is the visibility and $C$ is a normalization constant.
Without loss of generality we can assume that the efficiency of
the detectors is one.

However we do not access the time $t$ but we only measure $\Delta
t$. Therefore the probability of having two coincidences, knowing
only that the 2nd photons comes $\Delta t$ after the first is
given by
\begin{eqnarray}
&p\left( \Delta t|\gamma_{\Delta t}\right) =\frac{\Omega^0 }{2\pi }\int_{0}^{\frac{%
2\pi }{\Omega^0 }}p\left( t,t+\Delta t|\gamma _{t},\gamma
_{t+\Delta t}\right)
dt\\
&\cong C\frac{\Omega^0 }{2\pi }\int_{0}^{\frac{2\pi }{\Omega^0
}}\frac{1+V\cos \left( \Omega^0 t\right) }{2}\frac{1+V\cos \left(
\Omega^0 \left( t+\Delta t\right) \right) }{2}dt \nonumber
\end{eqnarray}
because typically $\tau _{d}<<\frac{2\pi }{\Omega^0 }$.

The integration gives
\begin{equation}
p\left( \Delta t|\gamma_{\Delta t}\right) =C\frac{1}{4}\left[
1+\frac{V^{2}}{2}\cos \left( \Omega^0 \Delta t\right) \right]
\end{equation}
Finally the probability density of having 2 coincidences
separated by a time $\Delta t$ is obtained using the fact that
the emission and detection are two Poissonian processes
independent of the beatings. Therefore the probability
$dp_{emmission}(\Delta t)$ of having two emissions separated by a
time $\Delta t$ is
\begin{equation}
dp_{emission}\left( \gamma_{\Delta t}\right) =\frac{1}{\tau }\exp
\left( -\Delta t/\tau \right) d\left( \Delta t\right)
\end{equation}
Hence the probability density $P(\Delta t)$ of having two
coincidence events separated by a time $\Delta t$ is given by
\begin{eqnarray}
dp\left( \Delta t\right) &=&p\left( \Delta t|\gamma_{\Delta
t}\right)dp_{emission}\left( \gamma_{\Delta
t}\right) \nonumber\\
&=&CP\left( \Delta t\right) d\left( \Delta t\right)
\end{eqnarray}
with $P\left( \Delta t\right) =\frac{1}{4}\left[
1+\frac{V^{2}}{2}\cos \left( \Omega^0 \Delta t\right) \right]
\frac{1}{\tau }\exp \left( -\Delta t/\tau \right) .$ We normalize
this expression such that
\begin{eqnarray}
1 &=&\int_{0}^{\infty }P\left( \Delta t\right) d\left( \Delta t\right) \nonumber\\
&=&C\frac{1}{8}\left( 2+\frac{V^{2}}{1+(\Omega^0) ^{2}\tau
^{2}}\right)
\end{eqnarray}
The final normalized expression is then
\begin{equation}
P\left( \Delta t\right) =\frac{\left[ 1+\frac{V^{2}}{2}\cos \left( \Omega^0 \Delta t\right) %
\right] \exp \left( -\Delta t/\tau \right) }{\tau \left( 1+\frac{V^{2}}{2}\frac{1}{%
1+(\Omega^0) ^{2}\tau ^{2}}\right) }
\end{equation}
Experimentally, we integrate $P\left( t\right) $ over a time bin
$t_{b}$ so the measured probability is
\begin{equation} p\left(\Delta t\right) =\int_{\Delta t}^{\Delta t+t_{b}/2}P_{N}\left(
t^{\prime }\right) dt^{\prime }
\end{equation}
The total counts $N_{c}$ in $m$ seconds in each time bin is
\begin{equation}
N_{c}\left(\Delta t\right) =\frac{m}{\tau }p\left(\Delta t\right)
\label{beatings_prob}
\end{equation}
The advantage of this method to see 2-photon beatings is that the
interferometer have only to be stable during the time between two
successive coincidences. In section III.F we present the results
of the beatings frequency using this method.

\section{Experimental setup}

\subsection{Principle}

The setup we use to test entanglement of the photon pairs with
frequency shift, is based on previous Franson-type experiments
\cite{TittelPRA}. The main conceptual difference is the frequency
shifts in one arm of the interferometers.

\subsection{Source}

The photon pairs are created by parametrical down-conversion in a
recently developed high efficient source. It is based on a
waveguide integrated on a Periodically Poled Lithium Niobate
(PPLN) substrate \cite{SebElLett}. Using a pump at 657 nm, it
generates degenerated photons at 1314 nm. We chose this wavelength
as it corresponds to a transparency window in optical fibers.
Hence it is possible to use this setup for long distance
transmission. An RG1000 \ filter is placed after the waveguide to
eliminate the pump light, and an additional interference filter
is used to narrow the generated photons bandwidth. The photon
pairs are coupled into a 50/50 fiber optics beam-splitter which
separate the twin photons.

Violation of Bell inequality has already been demonstrated with
this source \cite{PPLNQC}.

\subsection{Acousto-optic modulator as a moving beam-splitter}

As for Franson type experiments, we use Michelson interferometers
as analyzers. We replaced in each interferometer the
beam-splitters by AOM's (Acousto-Optic Modulators, Brimrose
AMF-100-1.3-2mm). They have two effects. First they induce a
frequency shift equal to the acoustic wave frequency as we will
see, second they can be seen as moving beamsplitter as required
for a relativistic Bell experiment \cite{as00}.

\begin{figure}[h]
\includegraphics[height=5cm]{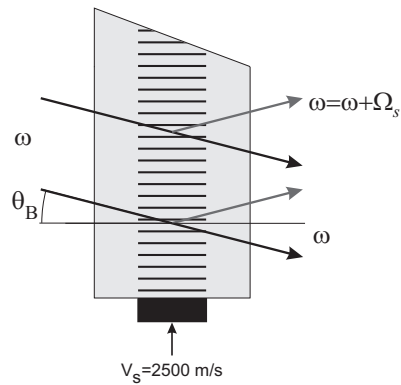}
\caption{Acousto-optic modulator, the Bragg grating created by a
sound wave of frequency $\Omega _{s}$ reflects part of the
incoming light, whose frequency is $\omega$} \label{AOM}
\end{figure}

An AOM is made of a piece of glass, AMTIR 1 (Amourphous Material
Transmitting IR Radiation), in which an acoustic wave at
frequency $\Omega $ (100 MHz in our experiment) is created by a
piezo-electrical transducer \cite{photonics}. As the refractive
index in a material depends on the pressure, the acoustic wave
will create a periodical change of the refractive index,
equivalent to a diffraction grating (fig. \ref{AOM}). If the
acoustic wave
is traveling rather than stationary, it will be equivalent to a \textit{%
moving} diffraction grating. This can be achieve if the AOM ends
with a skew cut termination to damp the wave. As for a standard
grating, the reflection coefficient is maximal at Bragg angle
$\theta _{B}$ given by
\begin{equation}
2\lambda _{s}\sin \theta _{B}=\lambda /n  \label{Bragg}
\end{equation}
where $\lambda _{s}$ is the sound wavelength, $\lambda $ is the
light wavelength in vacuum and $n$ the refractive index of the
material. The reflection coefficient is, for small angles
$\theta_B$ \cite{Yariv}
\begin{equation}
R=\frac{\pi ^{2}}{2\lambda ^{2}}\left( \frac{L}{\sin \theta_B }\right) ^{2}%
\mathcal{M}I
\end{equation}
where $I$ is the acoustic power, $L/\sin \theta_B $ is the
penetration of light through the acoustic wave, and $\mathcal{M}$
is a material parameter. The acoustic power can be set, such that
the beamsplitting ratio is 50/50.

The reflected wave undergoes a frequency shift of $+\Omega $ if
it hits the acoustic wave in the same direction as figure
\ref{AOM}, in the opposite case the frequency shift is $-\Omega $.

The reflection on a moving mirror produces a frequency change of
the light \cite{as00}, due to the Doppler effect, given by
\begin{equation}
\Delta \nu =\frac{2nv\sin \theta }{c}\nu  \label{Doppler}
\end{equation}
where $v$ is the mirror velocity and $\theta $ the angle between
the incident light and the plane of reflection. Within an AOM the
reflected light is also frequency shifted and the frequency shift
is equal to the acoustic wave frequency:
\begin{equation}
\Delta \nu =\Omega  \label{shift}
\end{equation}
Using $\lambda _{s}\Omega=v_{s}$, $\theta=\theta_{b}$ and
equations (\ref{Bragg}) and (\ref{shift}) we find that the
frequency shift induced by the AOM is the same as the one induced
by a mechanical grating traveling at speed $v_{s}$. The velocity
of sound in the AMTIR 1 can be computed from the mechanical
parameters of the material. The velocity for primary sound waves
\cite{sound_speed} is given by
\begin{equation}
v_{s}=\sqrt{\frac{\lambda +2\mu }{\rho }}
\end{equation}
where $\mu =\frac{E}{2\left( 1+\nu \right) }$ and $\lambda =\frac{E\nu }{%
\left( 1+\nu \right) \left( 1-2\nu \right) }$, $E$ is the Young modulus and $%
\nu $ is the Poisson ratio. For AMTIR 1 we have \cite{handbook}
$E=21.9\cdot
10^{9}$ Pa, $\nu =0.266$ and $\rho =4.41\cdot 10^{3}$ kg/m$^{3}$, hence $%
v_{s}=2480$ m/s. This corresponds to the manufacturer value of
$v_{s}=2500$ m/s.

The phase of the reflected wave is also shifted by a value
$\varphi$ which is the phase of the acoustic wave at the time when
the light is reflected.

\subsection{2-photon interferometry with frequency shifts}

\begin{figure}[h]
\includegraphics[width=8.43cm]{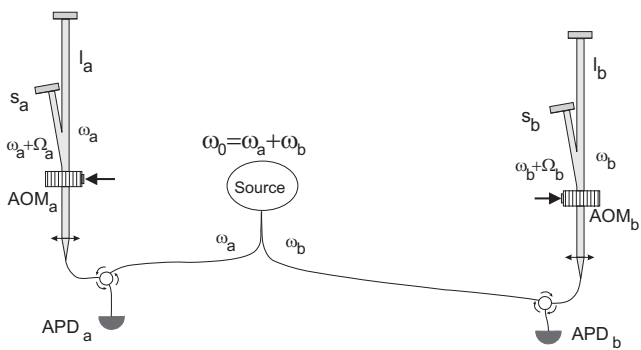}
\caption{Schematic of the experiment, the AOM's are oriented such
that both sound waves travel in opposite direction and such that
the frequency shift of the reflected wave are of opposite signs.
The total energy of both photons after the source is $\omega_0$.
When both photons pass through the long arms, the total energy is
not changed, but when they pass through the short arms the total
energy becomes $\omega_a+\omega_b+\Omega_a+\Omega_b$. It is
required that $\Omega_a+\Omega_b=0$ to avoid that beatings hide
the correlations.} \label{orientation}
\end{figure}

We have built two bulk Michelson interferometers using AOM's
instead of beamsplitters. The light is coupled out of the fiber
using an APC connector to avoid backreflection at the fiber's
end. Because of the small deviation
angle $2\theta_{b}$ (about 5$%
%TCIMACRO{\UNICODE[m]{0xb0}}%
%BeginExpansion
{{}^\circ}%
%EndExpansion
$) we collect only the light coming back into the input port by
using a fiber optical circulator (fig. \ref{orientation}). Due to
imperfect overlap of the modes, the transmission through each
interferometer is about 45\%, with monochromatic light. The
transmission through the reflected arm is reduced for large
bandwidth photons, because the deviation angle depends of the
light wavelength. Hence an AOM will act as a band-pass filter for
the reflected beam with a measured bandwidth of about 30 nm. To
minimize this effect which could reduce the fringes visibility we
have to insure that the bandwidth of the photons is smaller by
placing after the source a spectral filter (11 nm bandwidth).

\begin{figure}[h]
\includegraphics[width=8.43cm]{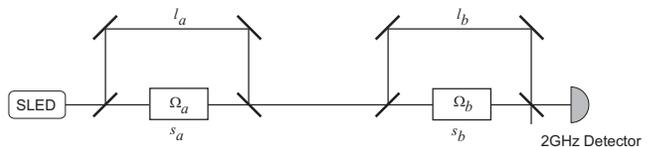}
\caption{Principle of alignment of the interferometers path
difference. Experimentally Michelson interferometers were used}
\label{alignement}
\end{figure}

The condition to observe interferences is given by equation
(\ref{Visibility}). It is the usual condition for 2-photon
correlations, the path differences $\Delta l_{i}$ between the
short and the long arm of the interferometers have to be equal
within the coherence length of the photons. Without frequency
shifts ($\Omega_a=\Omega_b=0$) the equalization of the paths of
both interferometers can be done by putting them in serie
\cite{TittelPRA} (fig. \ref{alignement}) and looking for the
interferences between the $s_a$$l_b$ and $l_a$$s_b$ paths. Using
a low coherence light, interferences appears only when $s_a+l_b$
is equal to $l_a+s_b$ within the coherence length of the light.
This implies that $\Delta l_a=\Delta l_b$ (in our interferometers
$\Delta l_i/c\simeq1.5 ns$). Unfortunately this method can not
directly work when there is a frequency shift. Indeed when we put
our interferometers in serie, with the AOM's oriented as for the
2-photon experiment, the frequency shifts do not cancel anymore
but beatings appear between the $s_a$$l_b$ and $l_a$$s_b$ paths
at the frequency of $\Omega_{a}-\Omega_{b}=400$ Mhz. In order to
observe those beatings we use a fast PIN detector (2 Ghz
bandwidth) connected to an electrical spectrum analyzer. As such
a detector is not very sensitive, we use a very bright LED source
(Opto Speed SA SLED 1300-D10A). With this setup it is easy to
scan the long path of one of the interferometers until we see
classical beatings.

Once the path length differences are equalized we can look for
2-photon interference with the setup of (fig. \ref{orientation}).
Therefore the frequency shifts have to cancel each other such that
$\Omega^{0}=0$ (eq. \ref{Coincidences rate with beatings}). We
orient the AOM as described such that the frequency shifts are of
opposite signs (fig. \ref{orientation}). Hence we have
$\Omega_a=200$ Mhz (because we pass 2 times through each AOM) and
$\Omega_b=-200$ Mhz.

Experimentally we can only specify an upper bound on
$|\Omega^{0}|$. The requirement is given by the fact that we have
to integrate over times much larger than $1/\Omega ^{0}$ to
estimate the coincidence probability with small statistical error
(typically 10-20 s). Therefore, even if the temporal resolution
of the detectors would have been good enough to see interference
fringes in principle (eq. \ref{VvsDt}), the integration cancels
them. Hence we can only see interference fringes if $\Omega
^{0}<10^{-2}$. Otherwise we have to use an indirect means of
observing the beatings, as described in the next section.

Finally one should note that if the frequency shift $\Omega$ is
greater than the photon bandwidth $\delta\omega_0$
\begin{equation}
\Omega\gg\delta\omega_0 \label{1-photon_distingu}
\end{equation}
it will be possible to distinguish the path by measuring only the
energy of one photon. In that case the correlations disappear
because the phase shift on the reflected wave will not be well
defined as the mirror is moving. More precisely this phase change
due to the change of the mirror position during the coherence
time $\delta\tau$ of the photon is given by
\begin{equation}
\Phi=\frac{\delta\tau
v_s}{\lambda_s}=\frac{\Omega}{\delta\omega_0}
\end{equation}
Interference fringes can be seen only if $\Phi\ll1$, which is in
contradiction with equation (\ref{1-photon_distingu}).

\subsection{Synchronization}

Each AOM is driven by a Radio-Frequency driver (Brimrose
FFF-100-B2-V0.8-E) which generates a 100 Mhz. As we have seen we
need to synchronize the Radio-Frequency drivers with a frequency
difference smaller than $10^{-2}$ Hz. This is done by using the
fact that the 100 MHz frequency is generated in each driver by
multiplying a 1 MHz signal from an oscillator with a phase locked
loop (PLL). The synchronization is achieved by using the same
oscillator for both drivers. In practice we send the signal from
one of the driver's oscillator to the other driver through a
coaxial cable. The
ratio of the frequencies, measured with a frequency-meter, is $%
1\pm10^{-11}$, so that $\Omega^0 <10^{-2}$ Hz. Another point to
look at is the shape of the electrical spectrum on both sides. We
verified with a spectrum analyzer that both spectrum widths are
smaller that the resolution of the analyzer (100 Hz).

\begin{figure}[h]
\includegraphics[width=8.43cm]{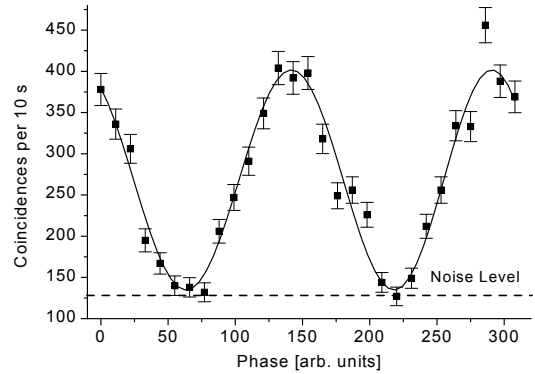}
\caption{Interference fringes. The dashed line indicates the
noise level, which is the rate of accidental coincidences. The
visibility after subtraction of the noise is $97\pm5\%$.}
\label{fringes}
\end{figure}

Once the synchronization is correctly done we observe interference
fringes with high visibility (fig. \ref{fringes}). The vivibility
after subtracting the accidental coincidences is about 97\%. The
visibility without subtracting the noise is about 45\%.

We have verified that the visibility is reduced when the
electrical spectrums of the drivers are slightly different. The
frequency of one of the drivers can be changed by steps of 15625
Hz. Due to beatings, no interference fringes can be observed when
the drivers are set at different frequencies.

\begin{figure}[h]
\includegraphics[width=8.43cm]{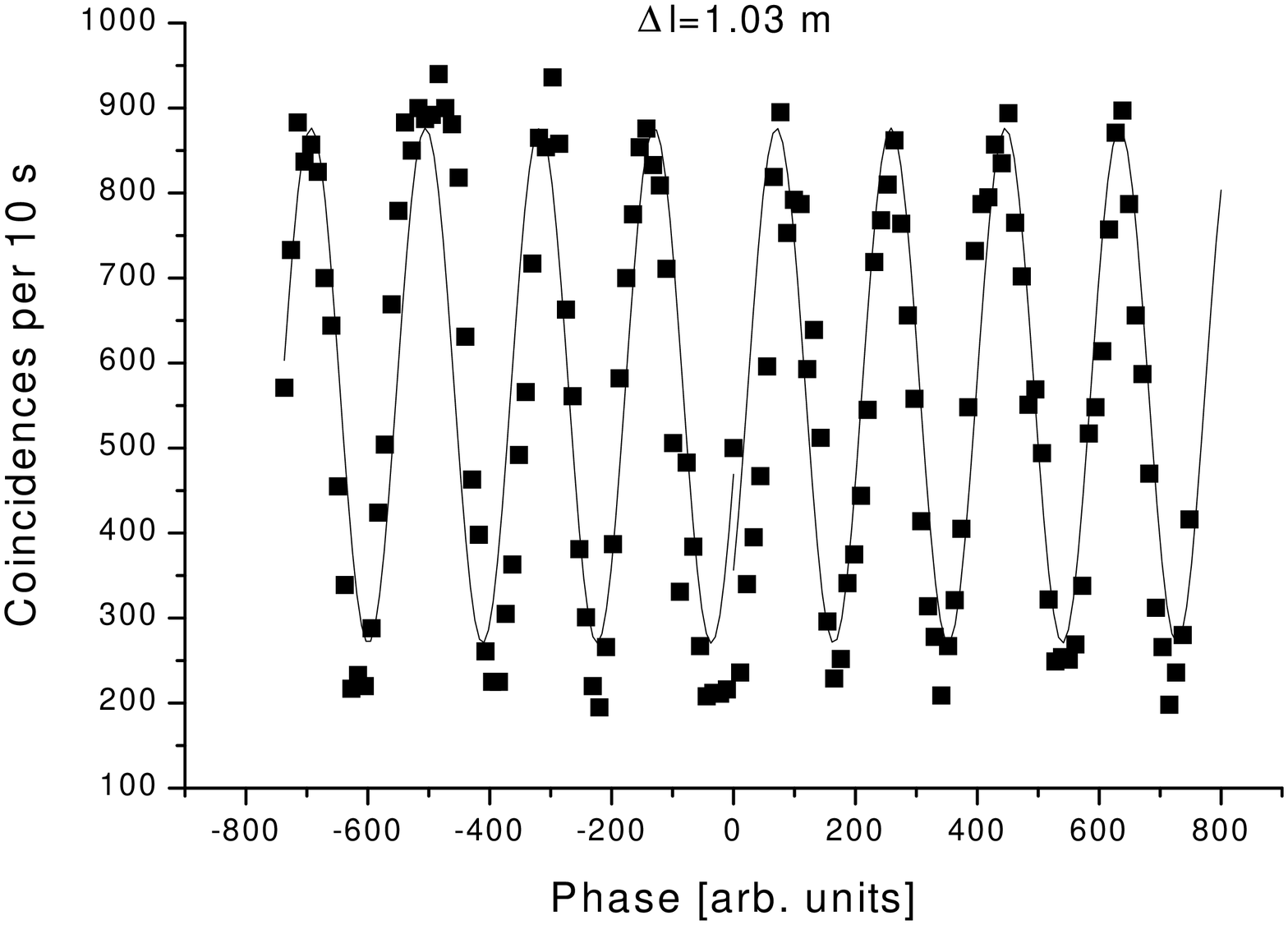}
\includegraphics[width=8.43cm]{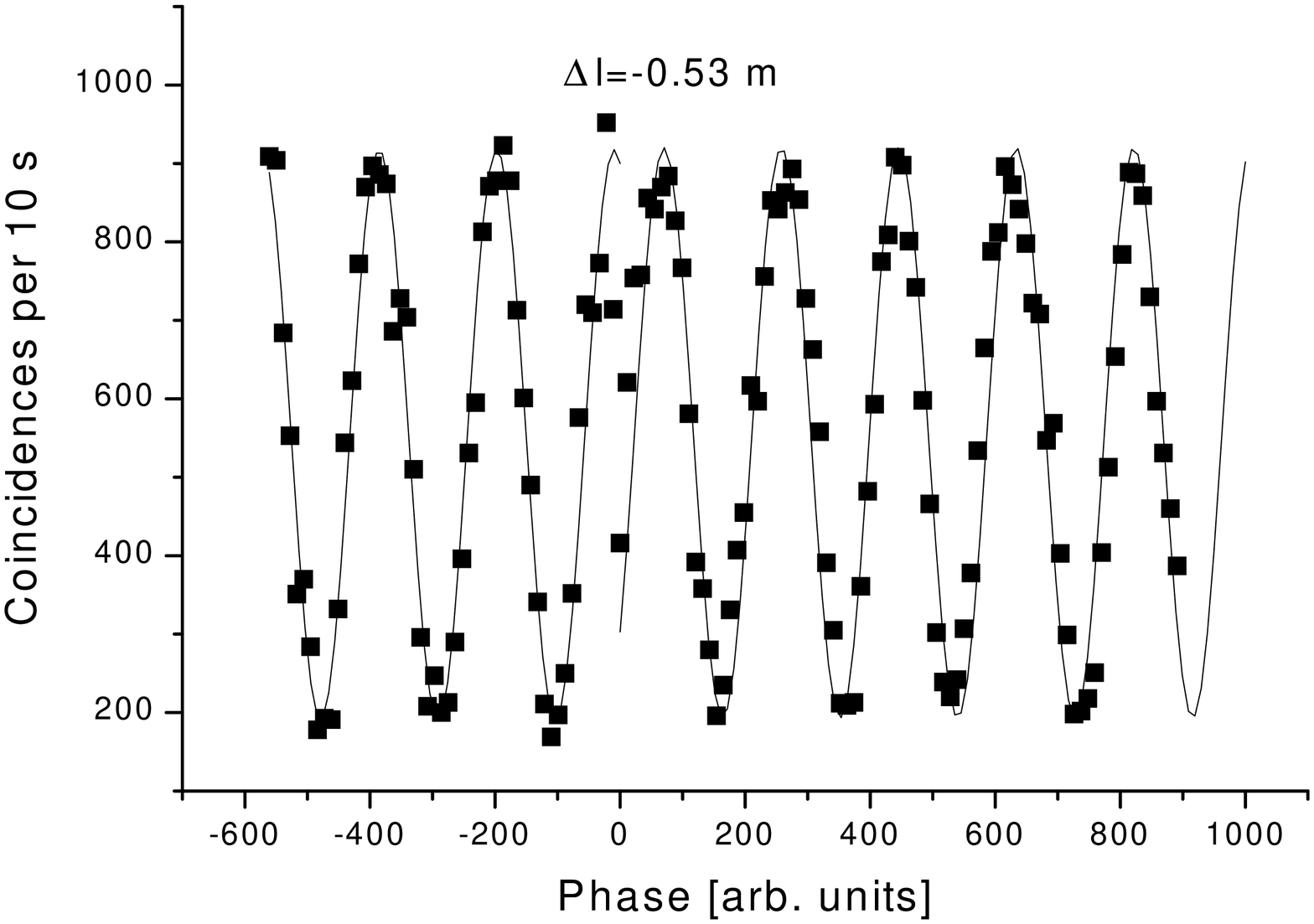}
\caption{Phase shift of the coincidence rate when the
synchronization cable length is changed.\newline Upper graph: 1.03
m is added, the phase shift given by the fit is
$\Delta\phi_{AOM}=6.73\pm0.20$ rad.\newline Lower graph: 0.53 m is
subtracted, the phase shift given by the fit is
$\Delta\phi_{AOM}=-3.61\pm0.07$ rad}.

\label{phase2pi}
\end{figure}

The phase difference $\phi_{AOM}$ between the 2 synchronized
acoustic waves induces the same phase change in the 2-photon
interference fringes. $\phi_{AOM}$ depends on the length $l$ of
the synchronization cable
\begin{equation}
\phi_{AOM}=\alpha l+\phi_0
\end{equation}
where
\begin{equation}
\alpha=\frac{2\pi}{\lambda_{synch}}\cdot\frac{\nu_{AOM}}{\nu_{synch}}
\end{equation}
with $\lambda_{synch}$ and $\nu_{synch}$ being the wavelength and
frequency of the 1 MHz synchronization signal. Hence we have
\begin{equation}
\alpha=\frac{2\pi \nu_{AOM}}{v_{synch}}
\end{equation}
with $v_{synch}$ the speed of the synchronization signal. We can
measure $\alpha$ by changing the length of the synchronization
cable and measuring the induced phase shift on the interference
fringes. We clearly observe a frequency shift when we add or
remove 0.53 m or 1.03 m of cable (fig. \ref{phase2pi}). The mean
phase shift per meter $\alpha$ of cable added on 5 measurements is
$6.97\pm0.09$ rad/m. Hence, with $\nu_{AOM}=2\cdot10^{8}$ Hz,
$v_{synch}=0.60c\pm0.01c$. This is compatible with the speed of
signal propagation in coaxial cables.

\subsection{Experimental evidence of 2-photon beatings}

\begin{figure}[h]
\includegraphics[width=8.43cm]{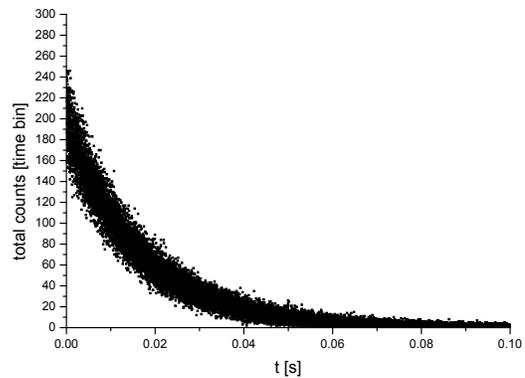}
\caption{Histogram of the time difference between successive
coincidences (large scale). This graph shows an exponential
decrease because the photon pairs detection is a Poissonian
process.} \label{exponentiel}
\end{figure}

\begin{figure}[h]
\includegraphics[width=8.43cm]{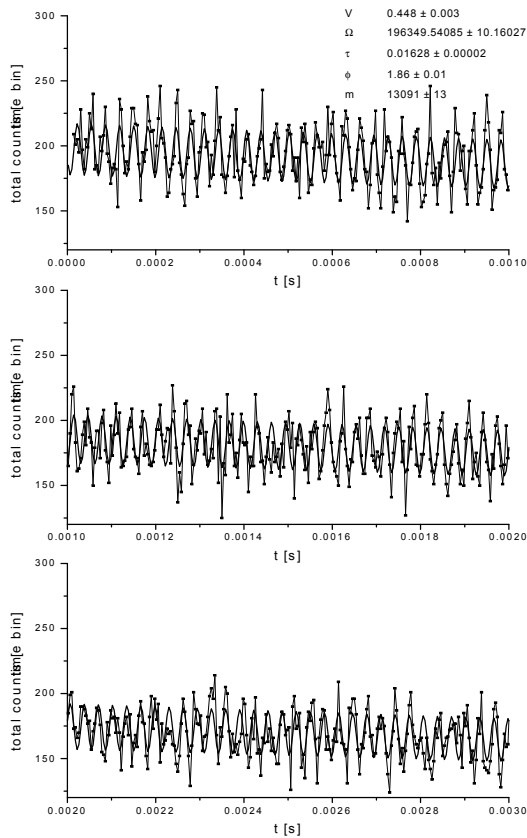}
\caption{Histogram of the time difference between successive
coincidences (small scale). The parameters of the fit of eq.
\ref{exp_beatings} are shown.} \label{beatings}
\end{figure}

We use the procedure described previously to experimentally show
2-photon beatings, when the difference of frequencies
$2\pi\Omega^{0}$ is 31250 Hz. We have measured the time
difference between successive coincidences and we plot the
histogram of those measurements. The time bins of the histogram
are of $4\cdot 10^{-6}$ s and we plot it for times between 0 s
and 0.1 s. The figure \ref{exponentiel} shows the exponential
decrease, as expected for random events. However at a closer look
we see that the exponential decay is modulated by a cosinus (fig.
\ref{beatings}). We fit an approximation of equation
(\ref{beatings_prob})
\begin{eqnarray}
N_{c}\left( t\right) &=&\frac{m}{\tau }p\left( t\right)\nonumber \\
&\cong &\frac{m}{\tau }t_{B}\frac{\left[ 1+\frac{V^{2}}{2}\cos
\left( \Omega
t\right) \right] \exp \left( -t/\tau \right) }{\tau \left( 1+\frac{V^{2}}{2}%
\frac{1}{1+\Omega ^{2}\tau ^{2}}\right) } \label{exp_beatings}
\end{eqnarray}
because the width of the histogram time bins is small enough (
$t_{B}<<1/\Omega $).

The visibility given by the fit is compatible with the direct
measurements of visibility and the frequency we found is
$\omega=31250.0\pm1.6$ Hz, as expected. Another measurement with a
frequency shift of 62.5 KHz gives similar results.

\section{Multisimultaneity}

\subsection{Motivation}

Classically, correlations between separated events can be
explained by two different mechanisms: either both events have a
common cause in the past, e.g. two separate TV apparatuses showing
the same images because connected to the same channel; or one
event has a direct influence on the other, e.g. dialing number on
my phone makes the phone of my colleague ring. In both cases there
is a time-ordered causal relation.

Quantum correlations, on the other hand, are of very different
nature. Quantum Mechanics predicts correlated outcomes in
space-like separated regions for experiments using pairs of
entangled particles. Many experiments have demonstrated such
quantum correlations, under several conditions
\cite{Tittel98,exp}, in perfect concordance with the quantum
mechanical predictions. The existence of correlations shows that
the outcomes of the two measurements are not independent. However,
in that case, violation of Bell's inequality rules out the common
cause explanation \cite {jb64}.

To explain the correlations one could invoke a possible influence
of one outcome on to the other. Since the correlated events lie in
space-like separated regions, such a direct influence would have
to be superluminal. Moreover this would define a preferred frame,
because the time-ordering between two space-like separated events
is not relativistically covariant.

One could imagine a unique preferred frame which is relevant for
all the quantum measurements. The pilot-wave model of de Broglie
and Bohm \cite{dbbh} assumes such a preferred frame. This model
perfectly reproduces the results of Quantum Mechanics, and the
assumed connections, though superluminal, cannot be used for
faster than light communication \cite{faster-light}. Moreover,
since Quantum Mechanics is independent of the timing, Bohm's
preferred frame is experimentally indistinguishable. Another
theory assuming a unique preferred frame has been proposed by
Eberhard \cite{Eberhart}. In this model the connection between
the events is not only superluminal but it propagates at a finite
speed, and leads to faster than light communication. If, in the
preferred frame, both choices occur in a time interval short
enough, the correlations would disappear as the influences would
not have the time to propagate. However, experimentally this
theory cannot be refuted, because the speed of the influence can
be arbitrary large and is not specified by the theory.

A different natural possibility would be to assume that the
relevant reference frame for each measurement is the inertial
frame of the massive apparatus, and to define a time-ordered
dependence by means of several preferred frames. This possibility
has been developed within a theory called Multisimultaneity
\cite{asvs97}. More specifically, Multisimultaneity assumes that
the relevant frame is determined by the analyzer's inertial frame
(e.g. a polarizer or a beam-splitter in our case). Paraphrasing
Bohr, one could say that the relevant frame, hence the relevant
time ordering, depends on the very condition of the experiment
\cite{Bohr}.

In Multisimultaneity, as in the pilot-wave model, each particle
emerging from a beam-splitter follows one (and only one) outgoing
mode, hence particles are always localized, although the guiding
wave (i.e. the usual quantum state $\psi $) follows all paths, in
accordance with the usual Schr\"{o}dinger equation. When all
beam-splitters are at relative rest, this model reduces to the
pilot-wave model and has thus precisely the same predictions as
quantum mechanics. However, when two beam-splitters move apart,
then there are two relevant reference frames, each defining a time
ordering, hence the name of Multisimultaneity. In such a
configuration it is possible to arrange the experiment in such a
way that each of the two beam-splitters in its own reference frame
analyzes a particle from an entangled pair before the other. Each
particle has then to ''decide'' where to go before its twin
particle makes its choice (even before the twin is forced to make
a choice). Multisimultaneity predicts that in such a
\textit{before-before} configuration, the correlations disappear,
in contrary to the quantum predictions.

Whereas quantum mechanics is nonlocal and independent of the time
ordering, Multisimultaneity assumes a nonlocal but time-ordered
dependence between the events. Nevertheless, this alternative
model is not in contradiction with any existing experimental data
prior to the present experiment. Furthermore, it has the nice
feature that it can be tested using existing technology. This
means that \emph{before-before} experiments are capable of acting
as standard of time-ordered nonlocality (much as Bell's
experiments act as standard of locality).

Since it would have been very difficult to put conventional
beamsplitters in motion, we used traveling acoustic waves as
beamsplitters to realize a \textit{before-before} configuration.
It has been argued that the state of motion of the moving acoustic
wave defines the rest frame of the beam-splitters \cite {as00}. We
would like to stress that a before-before experiment using
detectors in motion has already been performed confirming quantum
mechanics, i.e. the correlations didn't disappear
\cite{Zbinden01}.

\subsection{Experiment}

\begin{figure}[h]
\includegraphics[width=8.43cm]{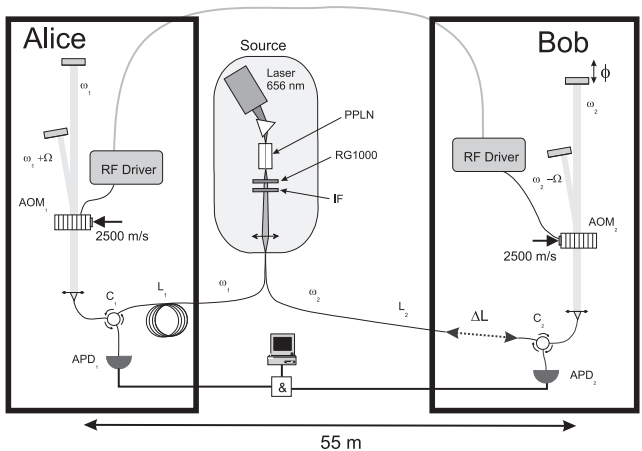}
\caption{Schematic of the experiment. The high efficiency photon
pair source uses a PPLN waveguide pumped by a 656 nm laser.
RG1000 filter is used to block the pump laser and a 11 nm large
interference filter (IF) narrows the photon bandwidth. The two
AOM's are 55 m apart and oriented such that the acoustic wave
propagate in opposite directions. One output of the
interferometers is collected back, thanks to optical circulators
$C_{1}$ and $C_{2}$ and the detection signals are send to a
coincidence circuit. As the frequency shifts are compensated, the
total energy when both photons take the short arms or the long
ones is constant. 2-photons interference fringes are observed by
scanning the phase $\protect\phi$} \label{exp_schem}
\end{figure}

\begin{figure}[h]
\includegraphics[width=6.00cm]{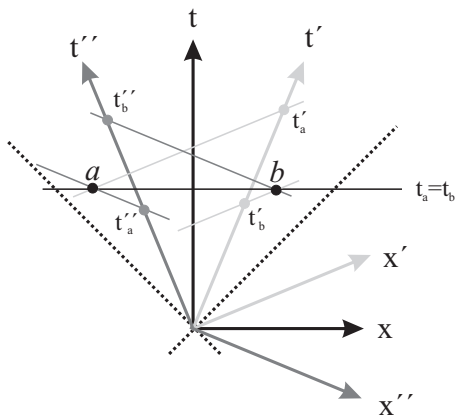}
\caption{The events $a$ and $b$ are simultaneous in reference
frame x-t, whereas $b$ is before $a$ in a frame x'-t' moving at
speed $v$ in x-t, and $a$ is before $b$ in frame x''-t'', which is
moving at speed $-v$ in x-t. The dashed lines represent the light
cone.} \label{frames}
\end{figure}

As we have seen, an AOM is a realization of a moving
beam-splitter. We can then use our interferometers to perform a
Bell experiment with moving beam-splitters (fig. \ref{exp_schem})
in order to confront quantum mechanics predictions with
multisimultaneity. We need to perform the experiment in the so
called \textit{before-before} condition. The criterion given by
special relativity for the change in time ordering of two events
in two reference frames counterpropagating at speed $v$ (fig.
\ref{frames}) reads
\begin{equation}
\left| \Delta t\right| <\frac{v}{c^{2}}d  \label{dt}
\end{equation}
where $\Delta t$ and $d$\ are respectively the time difference
and distance between the two events in the laboratory frame
\cite{asvs97}. This criterion is much more stringent than the
spacelike separation condition $\left| \Delta t\right|
<\frac{d}{c}$. Due to the high speed of the acoustic wave (2500
m/s), a distance of 55 m between the interferometers is enough,
and allows us to realize the experiment inside our building. The
permitted discrepancy on the time of arrival of the photons in
the AOM is then, according to (\ref{dt}), $\left| \Delta
t\right|_{max}=1.53$ ps, corresponding to a distance of $0.46$ mm
in air.

\subsubsection{Dispersion}

To observe the predicted disappearance of the correlations, the
spreading of the wave-packet due to the finite bandwidth of the
photons combined with the chromatic dispersion of the optical
fibers has to be smaller than $\Delta t$.

First, the coherence length of the single photons is determined by
the filter after the source. With a 11 nm filter the photons
coherence length is about 0.15 mm. Next, the chromatic dispersion
spreads the photon wave packet. However, thanks to the energy
correlation, the dispersion can be almost canceled. The
requirement for the 2-photon dispersion cancellation
\cite{Steinberg92} is that the center frequency of the 2 photon
is equal to the zero dispersion frequency of the fibers. We
measured this value on a 2km fiber with a commercial
apparatus (EG\&G) which uses the phase shift method. We found a value of $%
1313.2$ nm for $\lambda _{0}$. We use 100 m of the same fiber
assuming that the dispersion is constant along the fiber. We set
the laser wavelength at half this value. Knowing the chromatic
dispersion and conservatively assuming a 1 nm difference between
the laser wavelength and $\lambda _{0}/2$, the pulse spreading
over 100m can be computed and is 0.2 ps (for more detail see
\cite{Zbinden01}). This corresponds to a length of 0.06 mm in
air, which is much smaller than the permitted discrepancy. The total spread is given by $\sqrt{%
0.15+0.06^{2}}=0.152$ mm.

\subsubsection{Path length alignment}

\begin{figure}[h]
\includegraphics[width=8.43cm]{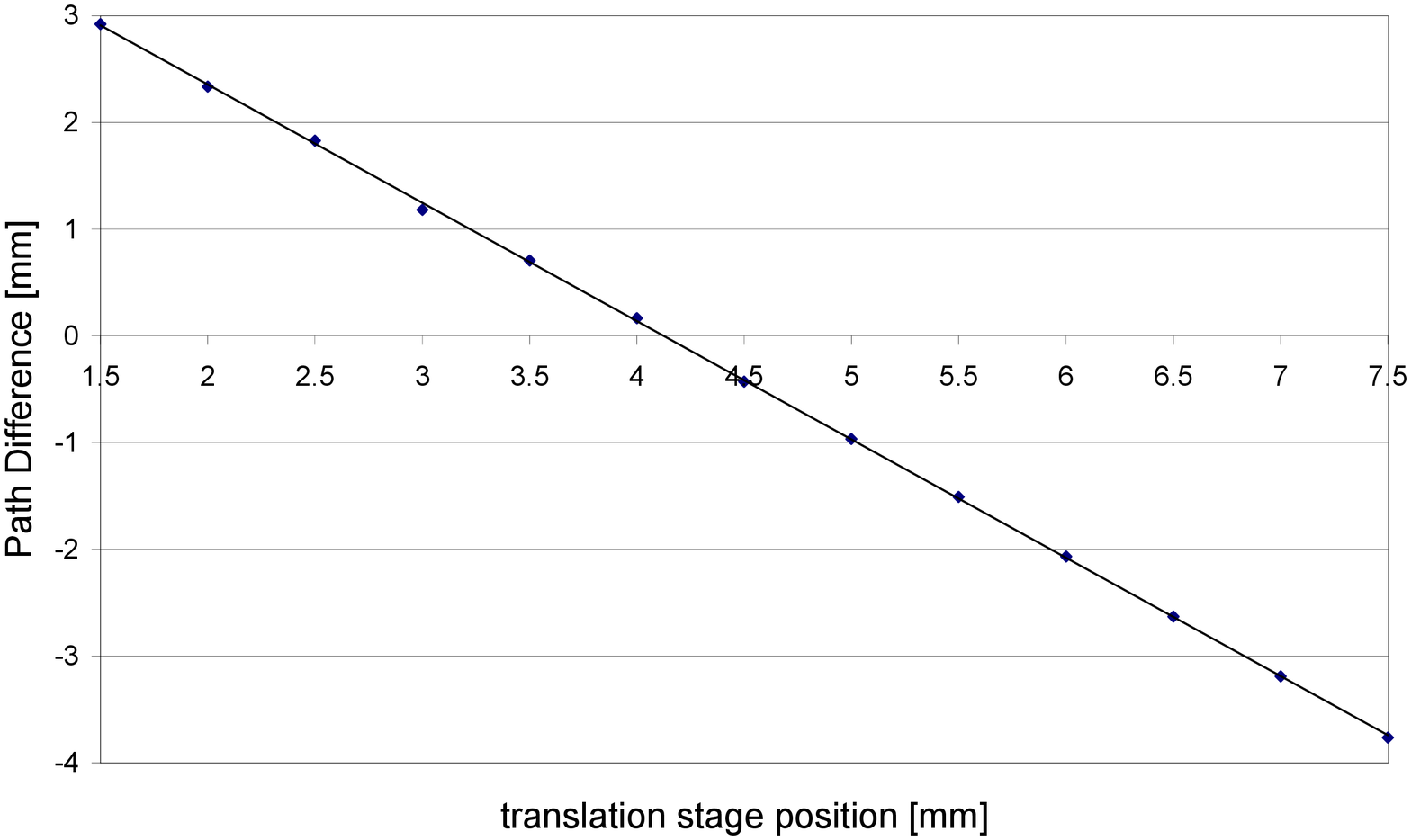}
\caption{Calibration of the optical path difference from the
$\Delta t=0$ situation vs. the position of the translation stage.}
 \label{calibration}
\end{figure}

The length difference between the fibers joining the source to two
the interferometers can be measured with a precision of 0.1 mm
using a low coherence interferometry method \cite{TittelPRA}. The
error on the interferometers' path lengths is measured manually
with a precision smaller than 0.5 mm. However, as the error is of
the order of $\left| \Delta t\right|_{max}$, we scan the path
length difference by pulling on a 1 m long fiber. One end of the
fiber is fixed on a rail as the other one is fixed to a
translation stage which is fixed to the rail. The fiber can be
elongated over a range of about 10 mm in the elastic region. As
the refractive index changes with the stress we need to calibrate
the effective change in the optical path vs. the fiber elongation
(fig. \ref{calibration}). We measure the visibility of the
interference fringes for each step of 0.10 mm corresponding to a
change of the path in air of 0.11 mm as given by the calibration.
This insure that for some of the scanning steps we are in the
\textit{before-before} situation.

\subsubsection{Results}

\begin{figure}[h]
\includegraphics[width=8.43cm]{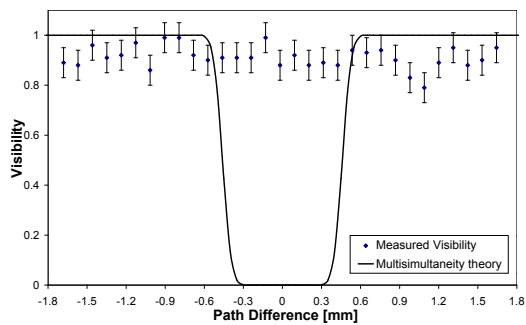}
\caption{Visibility vs path difference in the
\textit{before-before} situation. The dots are the measured
visibilities vs path difference by step of 110 $\mu$m. The
continuous line indicate the vanishing of the visibility as
predicted by multisimultaneity theory.} \label{VvsDelta}
\end{figure}

The theory predicts a disappearance of the correlations in the \textit{%
before-before} case. The visibility depending on the paths
difference would be given by the function

\begin{equation}
V=\left\{
\begin{array}{c}
{0\text{ if }\left| x\right| <\Delta t} \\
{1\text{ otherwise}}
\end{array}
\right.
\end{equation}

However as the photons have a non-zero coherence length and are
subject to spreading due to the dispersion, the correlations
would vanish smoothly. For a path difference of $x$ the
distribution of the times of arrival of the photons is given by
\begin{equation}
f\left( t\right) =\frac{1}{\sqrt{\pi }\sigma }\exp \left(
-\frac{\left( t-x/c\right) ^{2}}{2\sigma ^{2}}\right)
\end{equation}
Hence the visibility is given by the convolution of both
functions:
\begin{equation}
V_{true}(x)=\int_{-\infty }^{+\infty }\frac{1}{\sqrt{\pi }\sigma
}\exp \left( -\frac{\left( t-x/c\right) ^{2}}{2\sigma
^{2}}\right) V(t)dt
\end{equation}
In our case $\sigma =0.076$ mm due to dispersion and the photon
coherence length. The figure \ref{VvsDelta} shows the measured
visibility for different path difference and the expected curve
according to multisimultaneity. It is clear that there is no
disappearance of the correlations in the \textit{before-before}
situation.

Another intriguing situation is the opposite, where each
measurement device is analyzing its photon in its own reference
frame after the other analyzer photons. We call it
\textit{after-after} situation, for which Multisimultaneity also
makes predictions conflicting with Quantum Mechanics, and in our
particular case disappearance of the correlations, as in the
\textit{before-before} situation \cite{as97,as00}.

\begin{figure}[h]
\includegraphics[width=8.43cm]{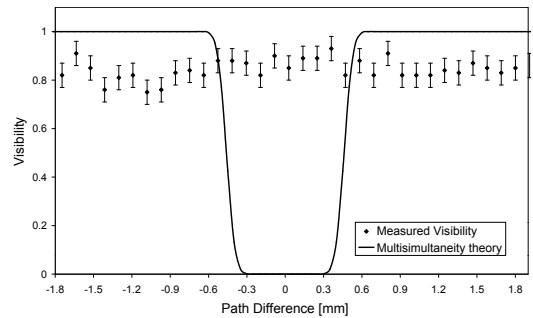}
\caption{Visibility vs path difference in the \textit{after-after}
situation.} \label{after}
\end{figure}

Experimentally, the \emph{after-after} situation is done by
inverting the direction of the acoustic waves, without changing
the other adjustments of the experiment. Figure \ref{after} shows
a measurement of the visibility in function of the path
difference. No change larger than the experimental fluctuations
can be observed around the 0 difference point.

\section{Conclusion}
We have modified the usual 2-photon Franson interferometry scheme
by using AOM's instead of standard beamsplitters. Adding a new
degree of freedom to the state of entangled photons leads to new
effects, in particular 2-photon beatings. These are equivalent for
the 2-photon interferences to the usual 1-photon beatings. We have
experimentally demonstrated 2-photon intereferences when the
beating frequency is 0. In the other case when it is not null we
have measured the beatings.

As the reflection on an acoustic wave is equivalent to the
reflection on a moving mirror we have used our interferometers to
test non-local correlations under different time orderings. In the
\emph{before-before} situation, each "choice-device" is the first
to analyze its photon in its own reference frame. In this
situation the correlation would disappear if  they where due to
some time-ordered influence between the events, as
Multisimultaneity assumes. Experimentally, we don't see any
vanishing of the correlations. Hence, not only, the quantum
correlations cannot be explained by local common causes as
demonstrated by violating Bell's inequality, but moreover one
cannot maintain any causal explanation in which an earlier event
influences a later one by arbitrarily fast communication.

In conclusion, correlations reveal somehow dependence between the
events. But regarding quantum correlations, our experiment shows
that this dependence does not correspond to any real time
ordering. Quantum entanglement cannot be cast into any
relativistic scheme.

\section*{Acknowledgments}

%==========================
This work would not have been possible without the financial
support of the ''Fondation Odier de psycho-physique'' and the
Swiss National Science Foundation. We thank Valerio Scarani and
Wolfgang Tittel for very stimulating discussions and Fran\c{c}ois
Cochet from Alcatel Cable Suisse SA for having placed at our
disposal the chromatic dispersion measurement instrument.

\appendix
\section{Frequency shift 2-photon quantum cryptography}
The quantum theory has allowed to develop new cryptographic
protocols, in particular quantum key distribution. Two people,
Alice and Bob, can create a shared secret key by exchanging
quantum particles. Coding bits into photons whose states are
randomly chosen between non-orthogonal bases prevent any effective
attack by a third person. A similar protocol based on the quantum
correlations between entangled particles has been proposed
\cite{Ekert91}. In this appendix we present two schemes for
quantum cryptography using 2-photon correlations. The first one
uses the phenomenon of
beatings as described in section 2, to simulate two basis at $45%
%TCIMACRO{\UNICODE{0xb0}}%
%BeginExpansion
{{}^\circ}%
%EndExpansion
$. The second is the analogous with 2 photon of the
interferometric scheme with frequency separation \cite{Merolla99}.

\subsection{Cryptography with pseudo-complementary basis}

\begin{figure}[h]
\includegraphics[width=8.43cm]{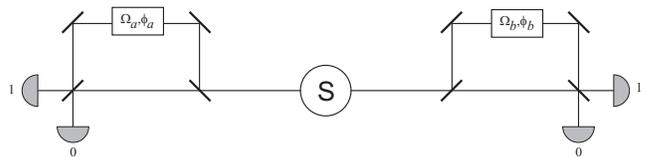}
\caption{Quantum key distribution scheme with entangled photon
pair. Depending on the phases and frequency shifts, correlations
can appear between the detectors outputs. If $\Omega_a+\Omega_b=0$
key distribution can be achieved by changing the phase $\phi_a$
and $\phi_b$, as if $\phi_a+\phi_b=0$ the frequency $\Omega_a$ and
$\Omega_b$ have to be changed.} \label{CryptoAOM}
\end{figure}

First, we briefly review the principle of 2-photon quantum
cryptography with phase coding \cite{Ekert91,crypto,Ribordy00}. It
is based on Franson 2-photon interferences (fig. \ref{CryptoAOM}).
Alice and Bob chose a phase in their respective interferometers
(corresponding to the choice of a basis). When the bases are
compatible ($\phi _{1}+\phi _{2}=0$) there is a perfect
correlation between the outputs of both interferometers, the
correlation coefficient $E$ is equal to 1. When the basis are
incompatible ($\phi _{1}+\phi _{2}=\pm\pi/2$) the correlation
coefficient vanishes, the outcomes are completly independant and
random. This is sumarized by the following table.

\[
\begin{tabular}{|l|l|l|l|}
\hline Alice $\phi _{1}$ & Bob $\phi _{2}$ & $\phi _{1}+\phi
_{2}$ & $E$
\\ \hline
$0$ & $0$ & $0$ & $1$ \\ \hline $0$ & $-\pi /2$ & $-\pi /2$ & $0$
\\ \hline $\pi /2$ & $0$ & $\pi /2$ & $0$ \\ \hline $\pi /2$ &
$-\pi /2$ & $0$ & $1$ \\ \hline
\end{tabular}
\]

Those correlations can be used to create a secret key between
Alice and Bob. This scheme could be implemented with our setup if
Alice and Bob set their frequency shifts such that $\Omega^0=0$.
They can change their respective phases by modulating the phase
in one arm (i.e. by changing the length). However they can also
change the phases by changing the phase of the synchronisation
signal on each sides as we have seen in section III E.

Instead of changing the phase Alice and Bob can change the
frequency of the photons. As they are looking for interferences
between the short-short and long-long paths, beatings will appear
if the sum of the frequency shift is not zero (section III F). The
probability coincidences are given by equation \ref {Coincidences
rate with beatings} where the global phase is fixed to $0$.
\begin{eqnarray}
p_{++} &=&p_{++}=\frac{1+\cos \left( \phi _{1}+\phi _{2}+(\Omega
_{1}+\Omega _{2})t\right) }{4} \\
p_{+-} &=&p_{-+}=\frac{1-\cos \left( \phi _{1}+\phi _{2}+(\Omega
_{1}+\Omega _{2})t\right) }{4}
\end{eqnarray}
Putting $\Omega _{1}+\Omega _{2}=0$ we find the previous table.
But we could change $\Omega _{1}+\Omega _{2}=\Omega^0 $ without
changing $\phi _{1}+\phi _{2}$. Hence if we set $\phi _{1}+\phi
_{2}=0$ we have
\begin{eqnarray}
p_{++} &=&p_{--}=\frac{1+\cos \left( \Omega t\right) }{4} \\
p_{+-} &=&p_{-+}=\frac{1-\cos \left( \Omega t\right) }{4}
\end{eqnarray}
If $\Omega^0 =0$ we find the perfectly correlated case, otherwise
they will be an additional phase of $\Omega^0 t$ which is not
equal to $\pm \pi /2$, as in the phase coding. However, as the
emission time of the photons is not known, the $\Omega^0 t$ value
is uniformly distributed,. So the mean value of $\cos \left(
\Omega^0 t\right) $ averaged over time is equal to $0$,
corresponding to the non correlated
case. We can then write a similar table as the previous one but with the phases $%
\phi _{i}$ replaced by the frequency shifts $\Omega _{i}$:
\[
\begin{tabular}{|l|l|l|l|}
\hline
Alice $\Omega _{1}$ & Bob $\Omega _{2}$ & $\Omega _{1}+\Omega _{2}$ & $%
\left\langle E\right\rangle $ \\ \hline $0$ & $0$ & $0$ & $1$ \\
\hline $0$ & $-\Omega $ & $-\Omega $ & $0$ \\ \hline $\Omega $ &
$0$ & $\Omega $ & $0$ \\ \hline $\Omega $ & $-\Omega $ & $0$ &
$1$ \\ \hline
\end{tabular}
\]
We should keep in mind that the correlation value in this table
is only a mean value, therefore we call it pseudo-complementary
basis. If Eve knows the detection time of a photon she could
follow the beatings, she will always know the value of $\cos
\left( \Omega^0
t\right) $ and can wait for a time $t_{Eve}=2\pi n/\Omega^0$, such that $%
p_{++}=p_{--}=1/2$, so to know which detector clicked on Alice
side. Therefore the detection time should be kept secret as long
as the photon didn't reach Bob side. However, Alice and Bob have
to perform a coincidence detection to select only the short-short
and long-long events. In order to discriminate these events, the
time uncertainty $\delta t$ on the coincidence has to be smaller
than the time difference $\Delta l/c$ between short and long arms.
Moreover, if the uncertainty on the detection time send by Alice
is greater than the beatings period, Eve can not extract any
information on the timing. This requires $1/2\pi\Omega^0 <\delta
t$. For example if $\Omega^0 =400$ Mhz we have $\delta t>0.4$ ns
so $\Delta l/c$ should be greater than 04 ns, which is the case
in our experiment ($\Delta l/c\simeq1.5$ ns).

Finally, if we imagine that Eve is able to measure when the photon
passes (by quantum non-demolition measurement), she can again
follow the beatings and set her measurement time such that she is
in perfect correlation with Alice. To avoid this attack Alice
should changed at random at which time she sent the photon to
Bob. Again, the uncertainty on the emission time $\delta t$ has
to be greater than $1/2\pi\Omega^0$.

Such scheme of quantum key distribution would not be easy to
implement, because it requires stabilisation of the
interferometers as for the phase coding. Moreover fast changes of
the frequencies of the AOM's are needed, but this is not the case
with today's modulators.

\subsection{2-photon quantum cryptography with phase modulation}

\begin{figure}[h]
\includegraphics[width=8.43cm]{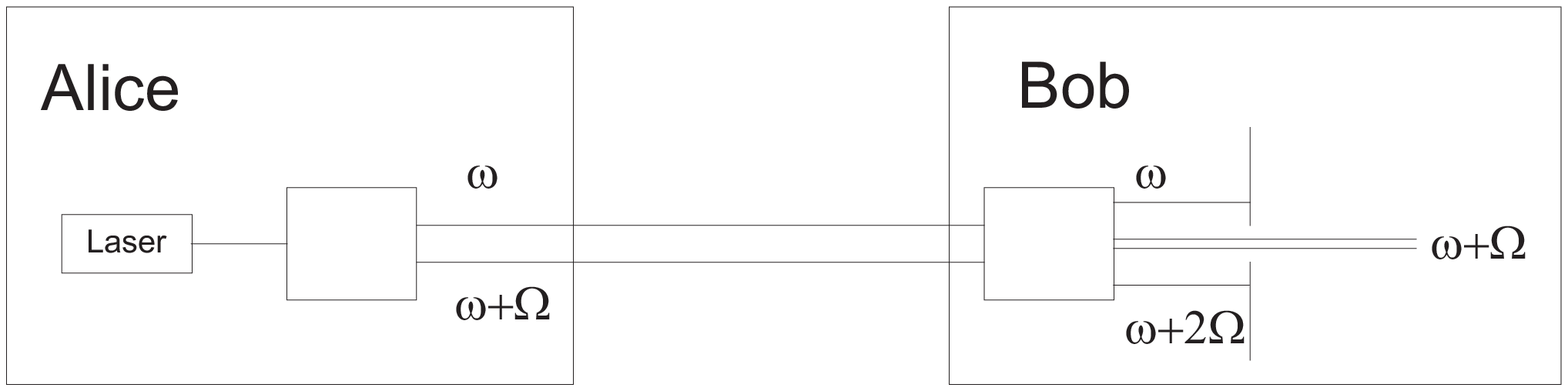}
\caption{Schematic of a quantum cryptography scheme using single
photon phase modulation.} \label{1photonBes}
\end{figure}

In contrary to the preceding scheme, where one photon
interferences occur between two spatially separated paths, a
scheme where the paths are frequency separated has been proposed
and realized in \cite{Merolla99,Duraffourg}. The experimental
setup was done using faint laser pulses, which, for this purpose,
is equivalent to a single photon scheme. It has the advantage of
not requiring an interferometers stabilization. We will see that
this one photon scheme can be generalized to a 2-photon one. But
first, we briefly review the original scheme. Alice creates a
photon in a superposition of two states of different frequencies
by modulating light at frequency $\Omega$ (fig \ref{1photonBes})
and by applying a phase difference $\varphi_a$. Hence the state
of a photon at frequency $\omega$ becomes

\begin{equation}
\left| \omega \right\rangle \rightarrow \left| \omega
\right\rangle +e^{i\varphi _{a}}\left| \omega +\Omega
\right\rangle \label{modulation}
\end{equation}
Bob analyzes the photon by applying a similiar operation with a
phase $\varphi_b$
\begin{eqnarray}
\left| \omega \right\rangle +e^{i\varphi _{a}}\left| \omega
+\Omega \right\rangle &\rightarrow &\left| \omega \right\rangle
+e^{i\varphi _{a}}\left| \omega +\Omega \right\rangle
+e^{i\varphi _{b}}\left| \omega
+\Omega \right\rangle \nonumber\\
&&+e^{i(\varphi _{a}+\varphi _{a})}\left| \omega +2\Omega
\right\rangle
\end{eqnarray}
and selecting only the $\omega +\Omega $ frequency. The remaining
state is
\begin{equation}
\left( e^{i\varphi _{a}}+e^{i\varphi _{b}}\right) \left| \omega
+\Omega \right\rangle \sim \left( 1+e^{i\left( \varphi
_{a}-\varphi _{b}\right) }\right) \left| \omega +\Omega
\right\rangle
\end{equation}

Such state can be used to implement quantum key distribution with
the B92 protocol \cite{Bennett92}.

\begin{figure}[h]
\includegraphics[width=8.43cm]{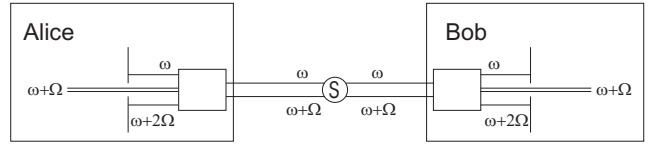}
\caption{Schematic of a quantum cryptography scheme using
2-photon phase modulation.}
\end{figure}

The generalization to 2-photon interferometry can be done because
each single photon cryptographic scheme is equivalent to a
2-photon scheme up to local unitary transformation \cite{crypto}.
In this case, the equivalent scheme consists in a source emitting
a state

\begin{equation}
\left| \omega +\Omega \right\rangle \left| \omega +\Omega
\right\rangle +\left| \omega \right\rangle \left| \omega
\right\rangle \label{source_state}
\end{equation}

When Alice and Bob apply the transformation given by eq.
(\ref{modulation}) one their side, the state becomes

\begin{widetext}
\begin{eqnarray}
\left| \omega +\Omega \right\rangle \left| \omega +\Omega
\right\rangle +\left| \omega \right\rangle \left| \omega
\right\rangle &\rightarrow &\left| \omega +\Omega \right\rangle
\left| \omega +\Omega \right\rangle +e^{i\varphi _{a}}\left|
\omega +2\Omega \right\rangle \left| \omega +\Omega \right\rangle
+e^{i\varphi _{b}}\left| \omega +\Omega \right\rangle \left|
\omega +2\Omega \right\rangle \nonumber\\
&&+e^{i\left( \varphi _{a}+\varphi _{b}\right) }\left| \omega
+2\Omega \right\rangle \left| \omega +2\Omega \right\rangle
+\left| \omega \right\rangle \left| \omega \right\rangle
+e^{i\varphi _{a}}\left| \omega +\Omega \right\rangle \left|
\omega \right\rangle \nonumber \\
&&+e^{i\varphi _{b}}\left| \omega \right\rangle \left| \omega
+\Omega \right\rangle +e^{i\left( \varphi _{a}+\varphi
_{b}\right) }\left| \omega +\Omega \right\rangle \left| \omega
+\Omega \right\rangle
\end{eqnarray}
\end{widetext}

In order to create the state of eq. (\ref{source_state}) by
downconversion, the laser pump has to be in the state
\begin{equation}
\left| \omega _{p}\right\rangle +\left| \omega _{p}+\Omega
/2\right\rangle
\end{equation}
It will generate the state

\begin{equation}
\int \left| \omega +\Omega +\delta \omega \right\rangle \left|
\omega +\Omega -\delta \omega \right\rangle +\left| \omega
_{p}+\delta \omega \right\rangle \left| \omega _{p}-\delta \omega
\right\rangle
\end{equation}
which doesn't need to be filter right after the pump, as the
photons with another energy than $\omega $ will not contribute to
the key exchange (they will be filtered on each sides).

After the filters the following state remains
\begin{equation}
\left( 1+e^{i\left( \varphi _{a}+\varphi _{b}\right) }\right)
\left| \omega +\Omega \right\rangle \left| \omega +\Omega
\right\rangle
\end{equation}
This state can be used to implement 2-photon quantum key
distribution. Moreover this scheme as only one output. It can be
generalized to a system with 2 outputs (BB84) by doing the
following frequency transformation
\begin{equation}
\left| \omega \right\rangle \rightarrow \left| \omega
\right\rangle +e^{i\varphi _{a}}\left| \omega +\Omega
\right\rangle +e^{-i\varphi _{a}}\left| \omega -\Omega
\right\rangle
\end{equation}

\end{document}